\documentclass[%
reprint,
superscriptaddress,
 amsmath,amssymb,
 aps,
 twocolumn,
]{revtex4-2}

\usepackage{graphicx}
\usepackage{dcolumn}
\usepackage{bm}
\usepackage{hyperref}
\usepackage{booktabs}
\usepackage{siunitx}
\usepackage{caption}
\usepackage{subcaption}
\usepackage{xcolor}
\usepackage{comment}
\usepackage[normalem]{ulem}
\usepackage[bottom]{footmisc}
\usepackage{cleveref}

\begin{document}

\preprint{}

\title{Investigations on the multi-sector hard X-Band Structures}


\author{V.A. Dolgashev}\thanks{Corresponding author. Email: dolgash@slac.stanford.edu}
\affiliation{SLAC, Menlo Park, CA, 94025, USA }

\author{L. Faillace}
\affiliation{INFN, Laboratori Nazionali di Frascati, P.O. Box 13, I-00044 Frascati, Italy}

\author{M. Migliorati}
\affiliation{Dipartimento di Scienze di Base e Applicate per l'Ingegneria (SBAI), Sapienza University of Rome, Rome, Italy}

\author{B. Spataro}
\affiliation{INFN, Laboratori Nazionali di Frascati, P.O. Box 13, I-00044 Frascati, Italy}

\date{\today}

\begin{abstract}

The development of ever advanced, high gradient accelerating structures is one of the leading activity of the accelerator community. In the technological research of new construction methods for these devices, high-power testing is a critical step for the verification of their viability.  Recent experiments showed that accelerating cavities made out of hard copper  fabricated without high-temperature processes, can achieve better performance as compared with soft copper ones. The results of experiments showed that welding, a robust and low-cost alternative to brazing or diffusion bonding, is an optimal solution for high-gradient operation, providing, with a gradient of about 150 MV/m, in the X-band, a breakdown rate of $10^{-3}$/pulse/meter using a shaped pulse with a 150 ns flat part. Within this framework, we are involved in the design, construction and higher power experimental tests of three cells standing-wave (SW) 11.424 GHz (X-band) accelerating cavities fabricated with hard Cu/Ag materials in order to study the RF breakdown physics. 
Our aim is to fabricate the accelerating structures with innovative technologies easier to handle and cheaper;  easier  for surfaces inspection; easier  for data elaboration and validation  of  joining techniques.   
The choice of these new technological approaches and design methods provides  also the possibility of  allocating the parasitic Higher Order Mode (HOM) dampers in case of regular multi-cells structure, both standing and traveling-wave accelerating structures, of advanced generation. This paper describes the design of an optimised cavity made with sectors that allows an increase of the frequency separation of longitudinal modes, and it provides a high longitudinal shunt impedance $R_{sh}$ of  the operating mode. The cavity will be fabricated by using the Tungsten Inert Gas (TIG) process in order to realise a hard Cu/Ag structure. Two three-cells  SW  X-band accelerating cavities, to be operated in the $\pi$-mode and made out of hard Cu/Ag alloy, were already fabricated at INFN-LNF by means of clamping and welding by using the TIG approach. Finally, we also report the RF characterisation and low-power RF tests of a two-halves split hard Cu/Ag structure that will be consequently TIG welded and employed for high-gradient tests and for the study of the RF breakdown physics.

\end{abstract}

\maketitle


\section{Introduction}

There is a strong demand for accelerating structures that are able to achieve high gradients for the next generation of linear particle accelerators for research, industrial, and medical applications. A continuous collaboration on the study of various geometries, materials, surface processing techniques, and technological developments of accelerating structures has involved, for more than a decade, the SLAC National Accelerator Laboratory in the USA, the Italian Institute of Nuclear Physics–National Laboratories of Frascati (INFN-LNF), and the High Energy Accelerator Research Organization (KEK) in Japan. This paper is part of this study and specifically focuses on the development of advanced accelerating structures aimed to show their feasibility by using new welding techniques {\color{red}\cite{ref3.1,ref3.2,ref3.3, ref3.4}}. To upgrade the performances of the X-band Linacs, and particularly to keep their length within reasonable limits, the accelerator community has concentrated resources and efforts to  study and develop ever more performing accelerator structures. The major interest is to achieve ever higher accelerating gradients, to increase the stored beam current and to keep a high reliability.
The development of ever more advanced accelerating structures {\color{red}\cite{ref0.1, ref0.2,ref27}} represents an important aspect in the activity of the accelerator community. High gradients, efficiency and good manufacturing of the accelerating structures play a fundamental role on linear accelerators. There is a strong demand of these structures able to achieve higher accelerating gradients and more compact dimensions for the next generation of linear accelerators. As an example, a XFEL photoinjector with a high field level larger than 250 MV/m, with a beam brightness increased by a factor of 25 increased beam brightness, would require about 1/3 the undulator length, and would have 3 times the photon beam peak power, as compared with existing FELs {\color{red}\cite{ref0.1, ref0.2}}. The linearizer is critically important for applications such as the MaRIE XFEL {\color{red}\cite{ref0.3}} , the CompactLight FEL \cite{ref26}, and the Ultra-Compact XFEL at UCLA {\color{red}\cite{ref27}}. Paired with the CompactLight sponsored initiative to develop a 15 MW-class klystron at 36 GHz, a compact, high gradient cryogenic linearizer in this frequency range now seems within reach {\color{red}\cite{ref0.4}}.

The superconducting cavities have reached performances that were unthinkable just a few years ago. New manufacturing techniques and sophisticated cleaning processes permitted to achieve, almost routinely, L-band RF accelerating gradients in the (25 - 30) MV/m range with some excellent results close to 45 MV/m {\color{red}\cite{ref1,ref2}}. 

However, it is very promising the copper coated with niobium in order to increase the radio-frequency properties. We sent to SLAC a few copper samples coated with 1 $\mu$m-thick niobium and a roughness of 20, 50 and 70 nm respectively,for checking the surface resistivity measurements {\color{red}\cite{ref2-3, ref2-3.1}}. 

The room-temperature accelerating structures too are being subject to intense Research and Development (R\&D) programs to reach higher accelerating gradients. In alternative to the wide diffused S-band and C-band, which provide an accelerating gradient of about 20 MV/m  and 50 MV/m {\color{red}\cite{ref3, ref4}}.

Respectively, the X-Band structures have been considered, since their reduced size and the high operating frequency may allow higher accelerating gradients.  As of today, stable operating gradients, exceeding 100 MV/m, have been demonstrated by the SLAC group in the X-Band (11.424 GHz). These experiments show that hard structures, fabricated without high-temperature processes, can achieve reliable accelerating gradients with RF breakdown rates better than those of soft annealed copper {\color{red}\cite{ref3.4,ref3.5,ref5}}. In  the framework of the collaboration, the involved laboratories decided to dedicate resources and manpower to study and develop hard advanced X-band accelerating structures to be used in the next generation projects.  

Two three-cells standing-wave accelerating structures, made out of different alloy materials, designed to operate in the $\pi$-mode at 11.424 GHz, have been successfully built and cold tested. The first structure is made out of OFHC copper and the second one is made out of Cu/Ag alloy with 0.08\% Ag concentration. Both cavities were fabricated at INFN-LNF and clamped together. In order to guarantee a vacuum envelope and mechanically robust assembly, we used the tungsten inert gas (TIG) process {\color{red}\cite{ref6,ref7,ref8}} for both structures.  

We are now involved in the optimised design as a function of the number of sectors (originally quadrants) {\color{red}\cite{ref3.4,ref3.5}}.

Due to experimental metallurgic results \cite{ref8.1,ref8.2},lower Ag concentrations in copper improve the surface conductivity therefore our goal is also to machine several samples with different lower concentrations than the current one in order to test the Rf performance. Moreover, higher Ag concentrations will be explored since the expected conductivity is also supposed to increase \cite{ref8.2} and as a final step we plan to realize copper alloys plated with Ag material, which has a higher conductivity than copper. The cavity will be consequently TIG welded and employed for high power tests and for the study of the RF breakdown physics. Our main interest is also to detune the cavity in order to reduce beam instability with a novel simpler technique and with dedicated absorbers of higher-order modes (HOMs).  

Multi-bunch instabilities are of major concern for high current machines. Intense beam currents and multi-bunch operation are essential features, for example, for increasing the luminosity of a linear collider, but multi-bunch beam break-up instability, which mostly arises from the parasitic modes of the accelerating structures, can limit the accelerator performance \cite{mosnier,ng,chao}. Also, in very high current circular machines, a longitudinal coupled bunch instability can arise from the cavity acceleration mode. To counteract these instabilities, an increase in the RF cavity bandwidth for the detuning frequencies is useful. To achieve this aim, the structure design made of quadrants that we propose (shown in Figure 1a, cavity design) allows getting a high longitudinal shunt impedance $R_{sh}$ of the accelerating mode, increases the mode separation frequencies. Moreover, we also  present the RF characterisation and low-power RF tests of one two-halves split hard-Cu/Ag structure. The two cavity halves are aligned and clamped together by means of male–female matching surface. The clamping is obtained with stainless screws, and the cavity will be TIG welded at COMEB srl company \cite{ref12}  on the outer surface. Low level-power RF measurements are in full agreement with the simulated ones. This structure will be consequently tested at high power for the study of the RF breakdown physics. This approach enables the construction of multi-cell standing and travelling-wave accelerating structure. 

Moreover, it is well known that waveguide (WG) can be used to extract HOMs power from the accelerating cavity as a means to cure multi-bunch instabilities. The WGs offer natural rejection to all the modes under cut-off, and therefore their use in HOMs extraction has the great advantage to trap the frequency of the operating mode whose frequency is below the WG cut-off frequency. The WGs were already proposed for HOMs damping in linear collider \cite{ref13}. The concept of using WGs in accelerating cavities has been stressed with the proposal of a 3-WG loaded cylindrical resonator which is free of HOMs \cite{ref14}. WGs  dampers have been adopted at SLAC for the PEP-II B-Factory project \cite{ref15}, for the accelerating cavity installed in the  DA$\Phi$NE machine \cite{ref16} and recently for the C-band linac in the framework of the Extreme Light Infrastructure-Nuclear Physics (ElI-NP) project \cite{ref17} . However such solution is also expensive.  

For the sake of completeness, in case of  high gradient accelerating structures,  the solution of WGs dampers is not suitable since they reduce the longitudinal shunt impedance $R_{sh}$ of the operating mode. For our applications, investigations on HOMs damping are also in  progress without adopting the WG dampers, but a simpler and cheaper approach.  Finally, the proposed TIG welded cavities are suitable for cryogenic operation and therefore they can be run at even higher gradients \cite{ref18,ref19}. A similar construction technique could be employed to build mm wave accelerating structures \cite{ref20,ref21}.

This manuscript describes the design of a Cu/Ag open accelerating structure as function of the sectors and discusses the preliminary low level RF measurements on a two halves structure in order to compare the theoretical estimations and the experimental ones. The cavity will be subject to high power tests in order to carry out studies of RF breakdown physics.

\section{Accelerating structure design criteria}

The design of  the particle accelerators of new generation is defined on the basis of a compromise among several factors: RF parameters, beam dynamics, RF power sources, easy fabrication, small sensitivity to construction errors, economical reasons and so on. In order to minimise the input power requirements for a given accelerating gradient, the RF accelerating structures have to be designed  with the aim of maximising the longitudinal shunt impedance $R_{sh}$ of the working mode. On the other hand, its performances could be limited by effects such as the beam loading, beam break-up etc., caused by the interaction between the beam and the surrounding environment.

\section{Beam Instabilities considerations}

It is well known that one of the main problems arising in the beam dynamics of high intensity accelerators is related to beam induced instabilities. By increasing the frequency separation of the accelerating structure passband allows to reduce or eliminate the instabilities associated to the operating accelerating mode and to achieve, at the same time, a high longitudinal shunt impedance $R_{sh}$.  The amount of detuning frequency can be estimated by \cite{ref22}:

\begin{equation}
    \frac{\Delta f}{f} = \frac{R_{sh}I_b \sin \phi_s}{2 Q V_c}
\end{equation}
where $V_c$ and $I_b$ are cavity voltage and beam current, respectively. The phase $\phi_s$ is the synchronous phase with respect to the crest of the RF voltage, and $Q$ is the quality factor. 

As a result, for a given beam current, gap voltage and geometry of the cavity, which provides the form factor \cite{ref22}:

\begin{equation}
    \frac{R_{sh}}{Q} = \frac{V_c^2}{\omega W L}
\end{equation} 
with $W$ the stored energy per unit lenght, $\omega$ the angular frequency, and $L$ the cavity length, we are able to estimate the frequency deviation. Structures loaded with an intense current  provide a large detuning. 

Since the structure under study is made of three cells, where the electric field of central cell is twice as high as that of the ends cells, the longitudinal shunt impedance calculation $R_{sh}$ was carried out only on the central cell, without taking into account the time transit factor of the cavity.  For the sake of completeness, the conclusions of this investigations can also be extended to the multi-cell or single-cell regular structure.

In the next section we investigate the frequency separation of the modes of the cavity's passband and optimise the Rf parameters of the working mode by modifying the cavity’s geometry.  

\section{Rf Structure Design} 

The choice of the cavity geometry has been matter of a long debate and related experimental activity in the framework of INFN-SLAC-KEK collaboration. The structure geometries have apertures, stored energy per cell, and rf pulse duration close to that of the NLC \cite{ref2-4,ref3} or CLIC \cite{ref9, ref10}.
The single-cell standing wave structure consists of three parts: the input coupler cell, the high-gradient middle cell, and the end cell \cite{ref3.2}. The geometry of the high-gradient middle cell is based on the geometry of a periodic accelerator structure cell. It has been found that the the behavior of the breakdown rate is reproducible for structures of the same geometry and material, and the breakdown rate dependence on peak magnetic fields is stronger than on peak surface electric fields for structures of different geometries \cite{ref3.3}. Main experimental tests of structures made from hard copper,, soft copper alloys and hard copper alloy are discussed in \cite{ref3.4}. The motivation for the study of  hard copper and the hard and soft-copper alloys came from results of our pulse heating experiments \cite{ref3.5}.

Here we present the detailed design strategy of a hard three cells open SW normal conducting accelerating structure in X-band as function of the number of sectors \cite{ref11} in order to maximise the accelerating gradient of the working mode for studying the Rf breakdowns physics and in the meantime investigate the modes separation frequency for reducing or eliminating instabilities associated to the operation mode. 

The best design is characterised by a simple geometry which is easy and cheap to construct with reasonable tolerances \cite{ref11}. The shape has been chosen in order to fabricate a hard accelerating structure by using the TIG method, achieving, at the same time, a high accelerating gradient and a large separation of the modes frequency. The structure is made of Cu/Ag alloy for increasing the breakdown threshold. This alloy has a satisfactory electrical and thermal conductivity to reduce the RF losses and to ease the cooling. 

The simulation studies have been performed as function of the number of sectors by using ANSYS HFSS software \cite{ref23}. Figure \ref{FullCavityE} and \ref{Full_Cavity_H} show the electric and magnetic field distributions, respectively, on the symmetry-planes of the full cavity which are scaled to a 100 MV/m accelerating gradient in the
middle cell.

\begin{figure}[h!]
\centering
\includegraphics[width=0.7\columnwidth]{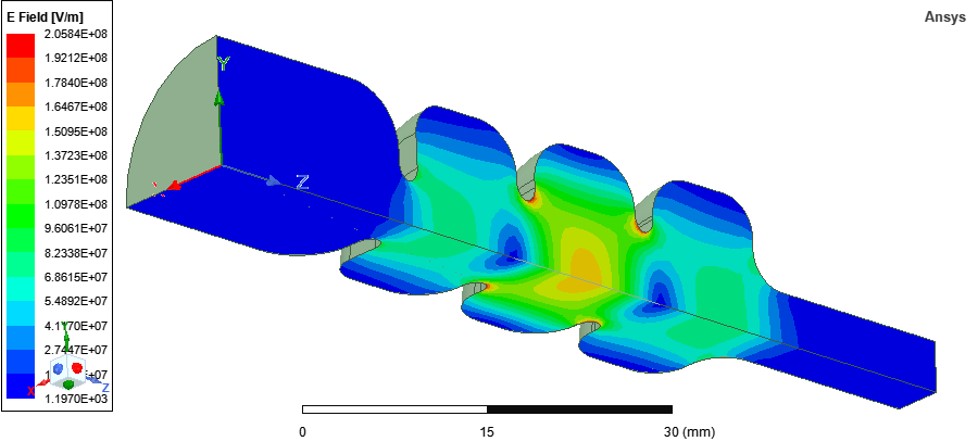}
\caption{\label{$Full_Cavity_E}Electric field distribution on the symmetry-planes of the full cavity which are scaled to a 100 MV/m accelerating gradient in the
middle cell. The surface electric maximum field is 205.8 MV/m.}
\end{figure}

\begin{figure}[ht]
\centering
\includegraphics[width=0.7\columnwidth]{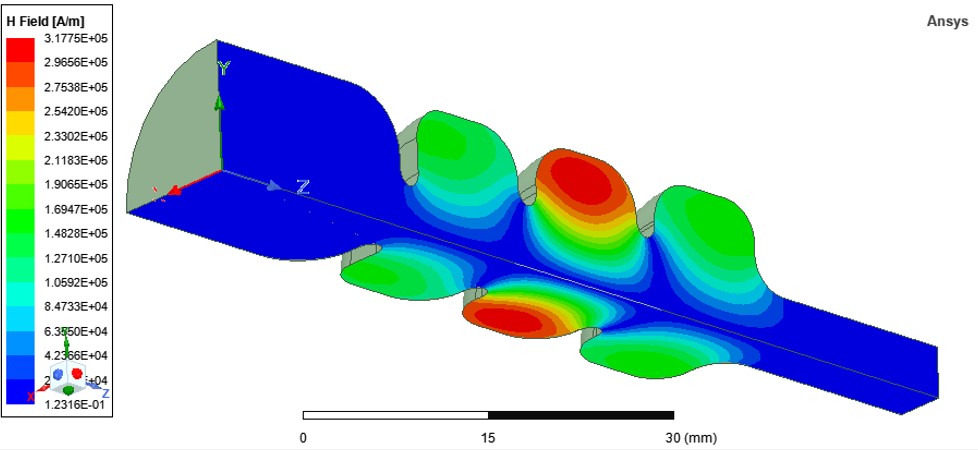}
\caption{\label{Full_Cavity_H}Magnetic field distribution on the symmetry-planes of the full cavity which are scaled to a 100 MV/m accelerating gradient in the
middle cell. The surface magnetic maximum field is 317.7 kA/m.}
\end{figure}

\begin{figure}[ht]
\centering
\includegraphics[width=1\columnwidth]{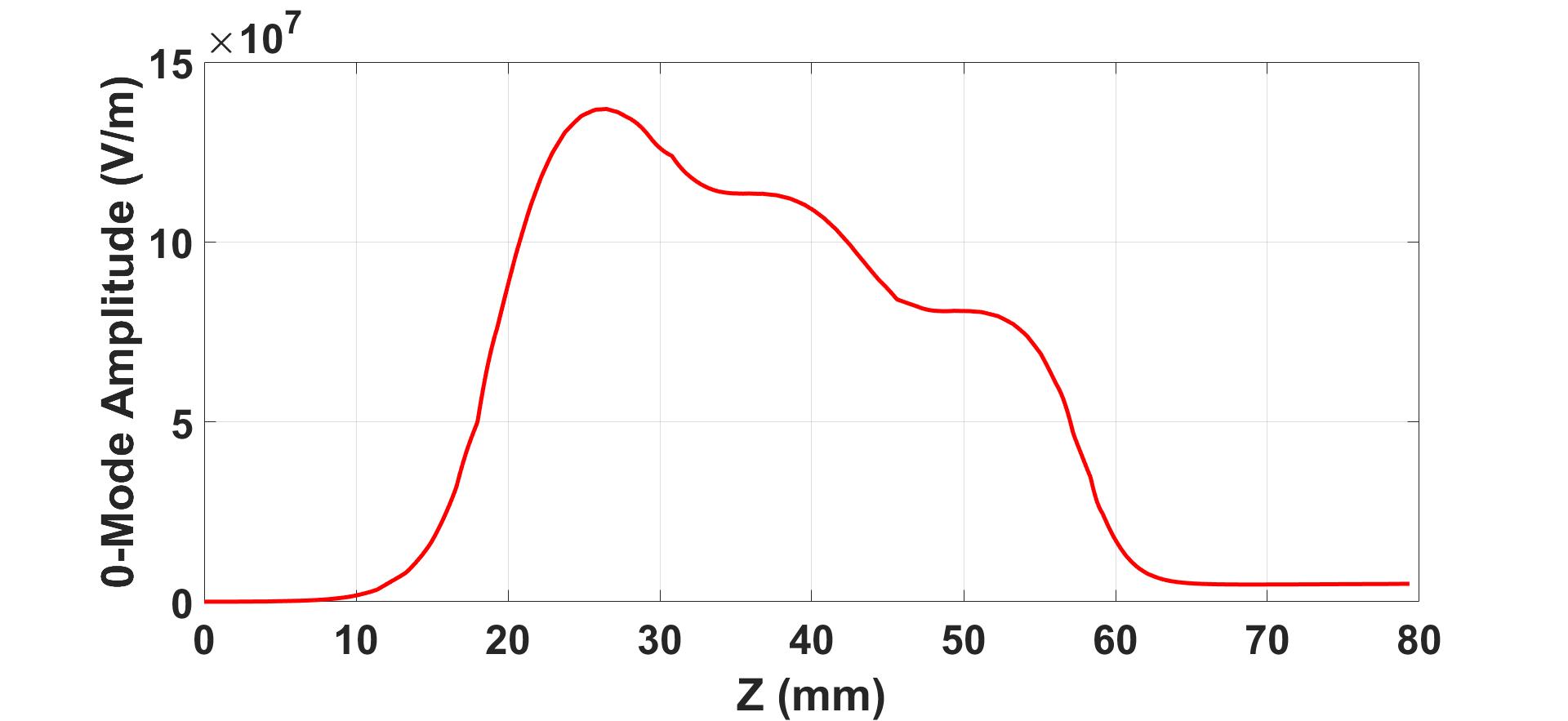}
\caption{\label{Zero_mode}Electric field profile of the 0 mode}
\end{figure}

\begin{figure}[ht!]
\centering
\includegraphics[width=1\columnwidth]{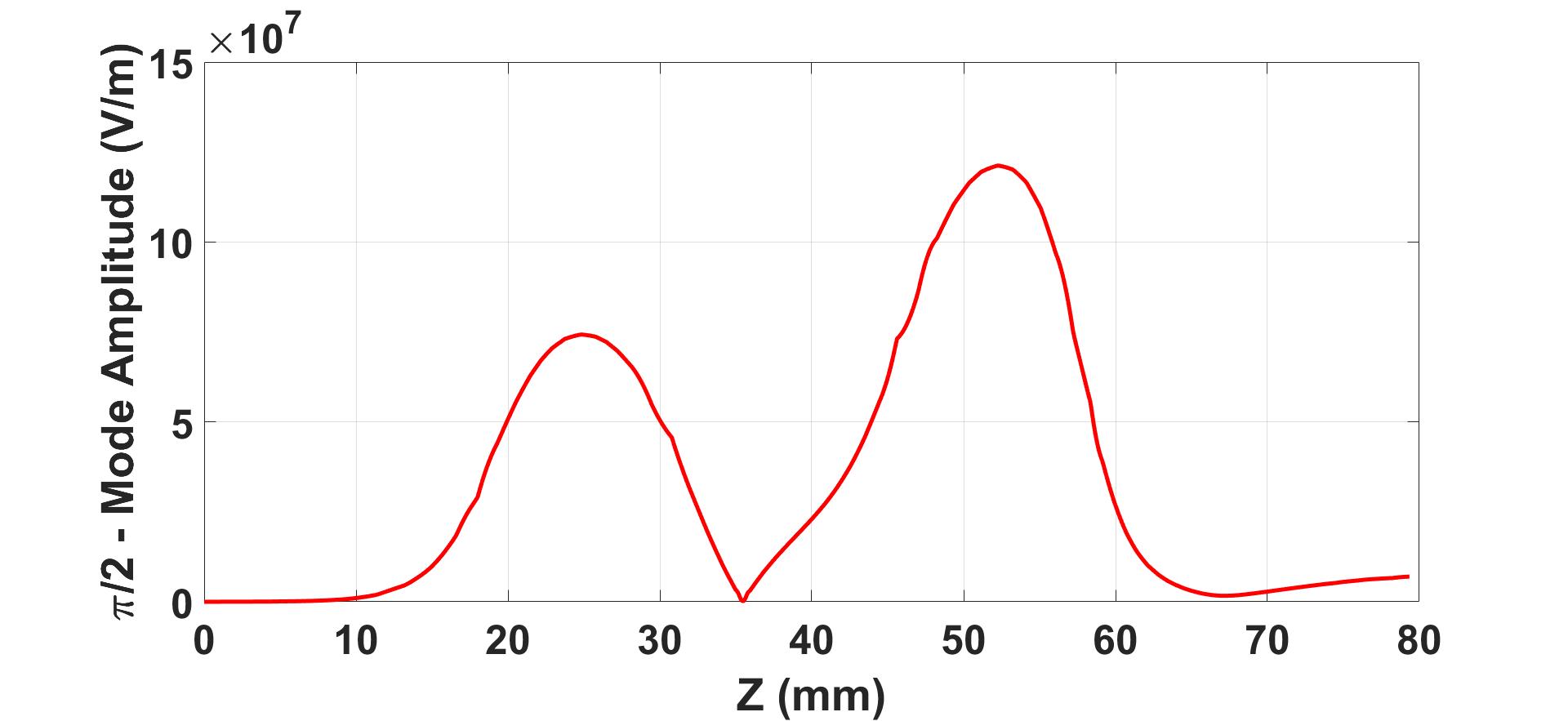}
\caption{\label{Pi2_mode}Electric field profile of the $\pi/2$ mode}
\end{figure}

\begin{figure}[ht!]
\centering
\includegraphics[width=1\columnwidth]{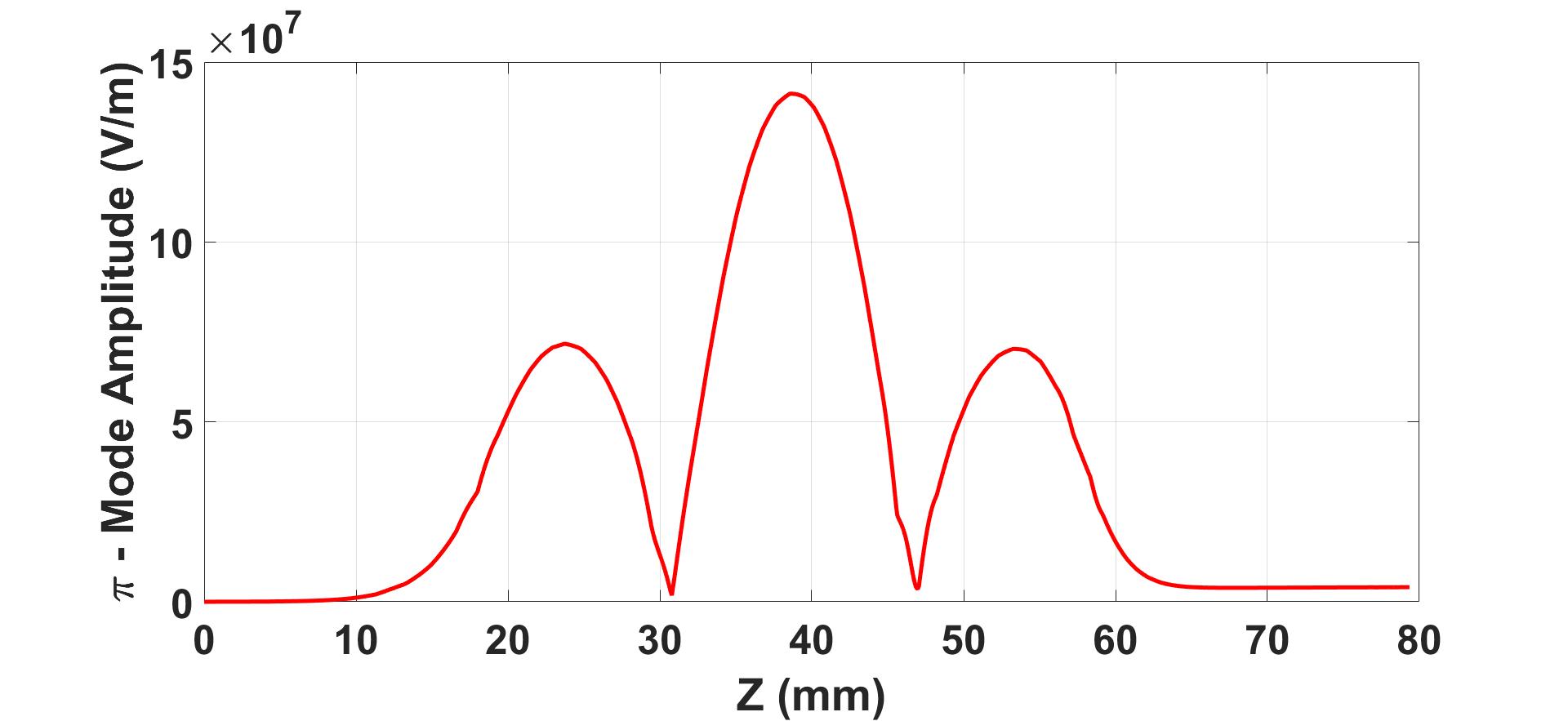}
\caption{\label{Pi_mode}Electric field profile of the $\pi$ mode}
\end{figure}

Figures \ref{Zero_mode}, \ref{Pi2_mode} and \ref{Pi_mode} show the  longitudinal field profiles of cavity modes 0, $\pi/2$ and $\pi$, respectively. 

The estimation of the longitudinal shunt impedance $R_{sh}$ of the operating $\pi$ mode has been carried only on the central cell of the structure by not taking into account the beam transit time factor.  In table \ref{tab:Cavity_RF_Parameters}, we report, as function of the number of sectors, the RF parameters of the operating $\pi$ mode: the quality factor $Q$, shunt impedance $R_{sh}$, the surface electric field $E_{surf}$, and the modified Poynting vector $S_c$ normalised at 100 MV/m. It can be noticed that the surface electric field enhancement is three times higher with respect to that of the full structure case. This value, around $E_{surf}=300$ MV/m, is due to the edge, properly rounded, between the cavity profile and the cut. For the same reason, the modified Poynting vector is amplified to about $S_c=5.7 MW/mm^2$ (slightly less than a factor 2) with respect to the full structure case. However, we are below the upper limit of breakdown threshold, as estimated by \cite{ref10}. Nevertheless, it is important to note that the values of all surface fields quantities are roughly independent of the number of slots.

\begin{table}[h]
\caption{\label{tab:Cavity_RF_Parameters}
 main Rf parameters as function of the number of sectors}
\begin{ruledtabular}
\begin{tabular}{ccccc}
Parts  & Q &$R_{sh} [M\Omega/m]$ & $E_{surf} [MV/m]$ & Modified \\

Number&&&&$S_c$\\
\hline
0 & 10608 & 138  & 231   & 3.3  \\
2 & 10520 & 136  & 292   & 5.6   \\
3 & 10476 & 136  & 286   & 5.8   \\
4 & 10432 & 135  & 285   & 5.7   \\
6 & 10343 & 133  & 300   & 5.6    \\
8 & 10253 & 131  & 300   & 5.7    \\
\end{tabular}
\end{ruledtabular}
\end{table}

Figure \ref{Modes_separation} shows, as function of the number of sectors $n$, the frequency variation of the modes 0, $\pi/2$ and $\pi$.  From inspection of the figure, we observe that,  by increasing the number of splittings, it is possible to increase the mode frequency separation accordingly, since the capacitive effect becomes stronger in the gap regions.

\begin{figure}[ht]
\centering
\includegraphics[width=0.8\columnwidth]{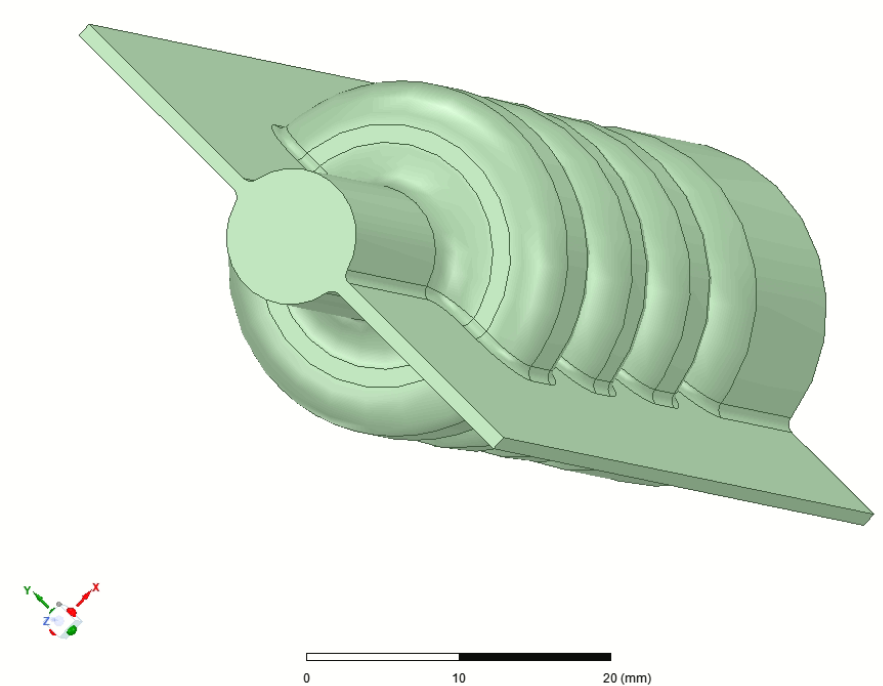}
\caption{\label{Halves_CAD}Surface model of the two-halves structure from HFSS.}
\end{figure}

\begin{figure}[ht]
\centering
\includegraphics[width=1\columnwidth]{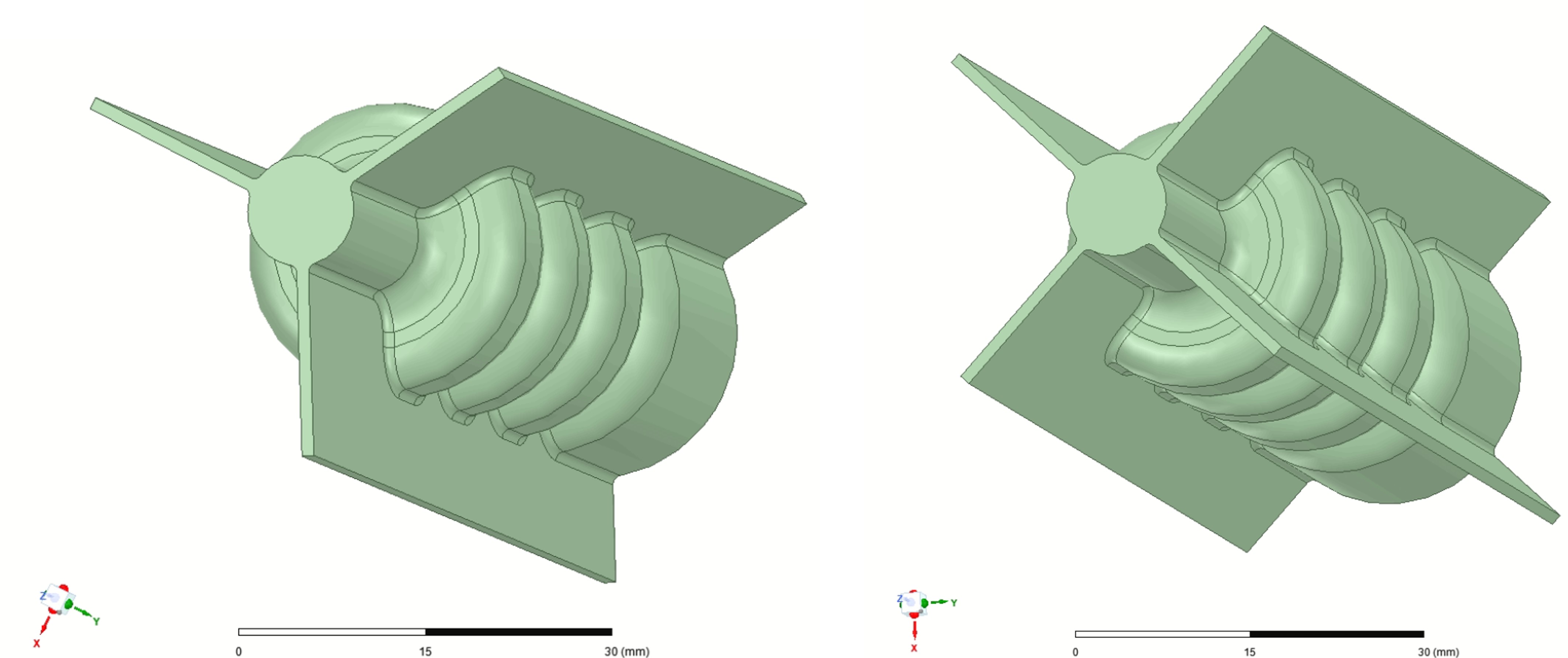}
\caption{\label{Three_Quads_CAD}Surfaces models of the three-parts structure (left) and four-parts or quadrants structure (right) from HFSS.}
\end{figure}

\begin{figure}[ht!]
\centering
\includegraphics[width=1\columnwidth]{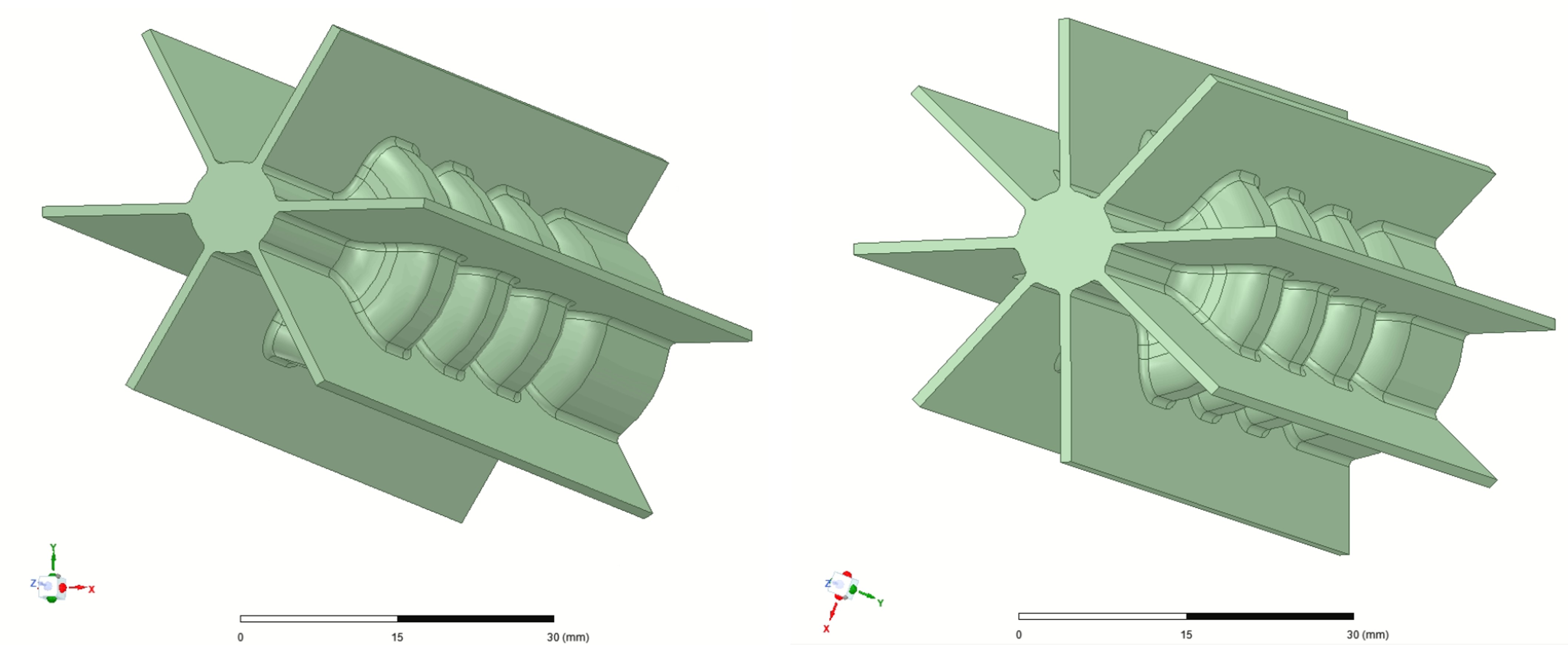}
\caption{\label{Six_Octants_CAD}Surfaces models of the six-parts structure (left) and eight-parts or octants structure (right) from HFSS.}
\end{figure}

\begin{figure}[ht]
\centering
\includegraphics[width=1.1\columnwidth]{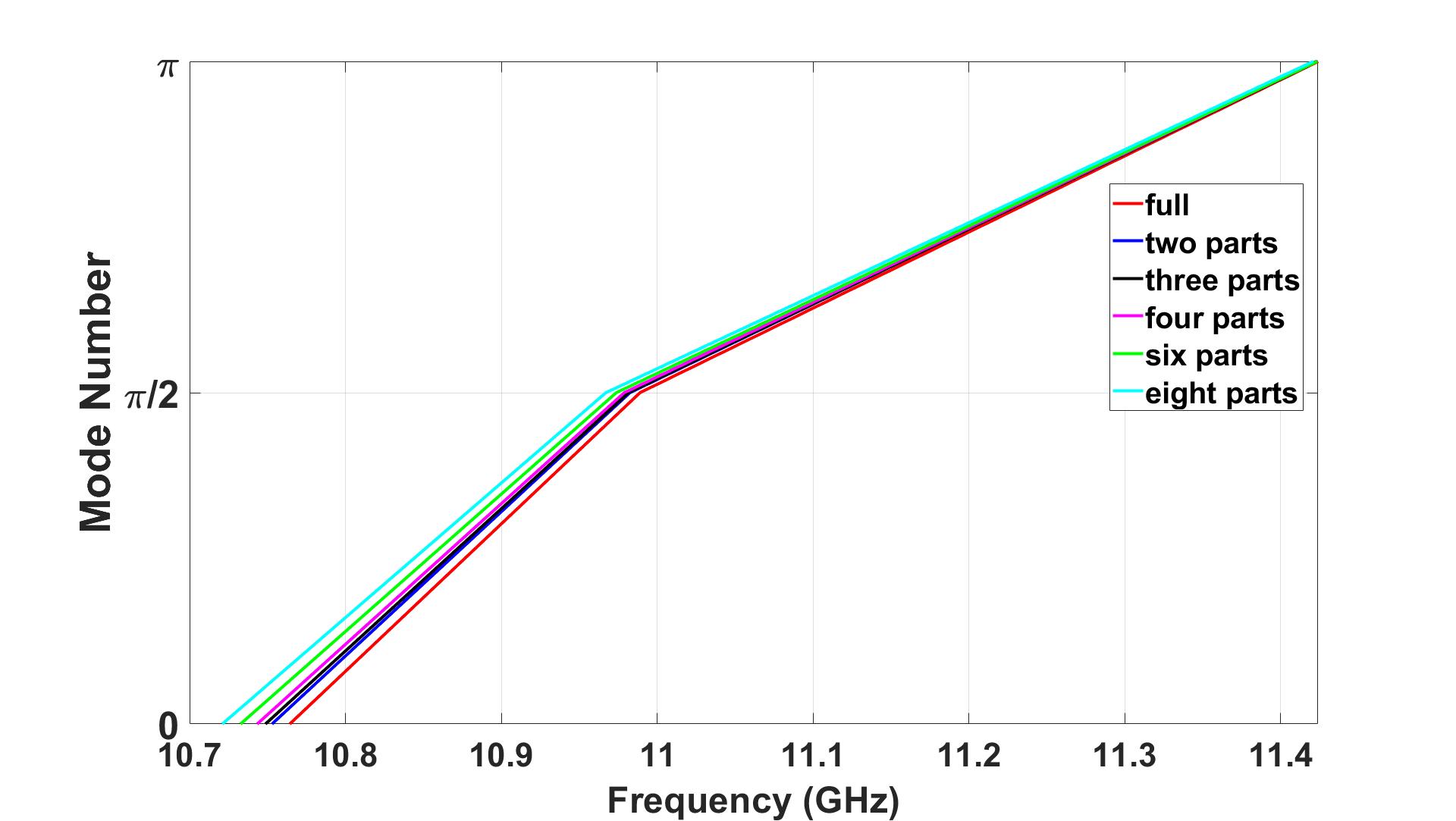}
\caption{\label{Modes_separation}Longitudinal Mode frequencies separation for all parts numbers.}
\end{figure}

The resonant frequency of the operating $\pi$ mode is kept constant at $F=11.424$ GHz for all splitting numbers, by properly adjusting each cell radius. The  frequency change rate of the 0 mode is $dF/dn  \simeq  5.41$ MHz, of the $\pi/2$ mode is $dF/dn \simeq  2.73$ MHz. Assuming 8 quadrants, we get about a 43 MHz frequency separation for the 0 mode, 21 MHz for the $\pi/2$ mode. As a result, this design method  improves the detuning effect of the 0 and $\pi/2$ modes of the structure without affecting the frequency of the working $\pi$ mode.

\begin{figure}[ht]
\centering
\includegraphics[width=1.1\columnwidth]{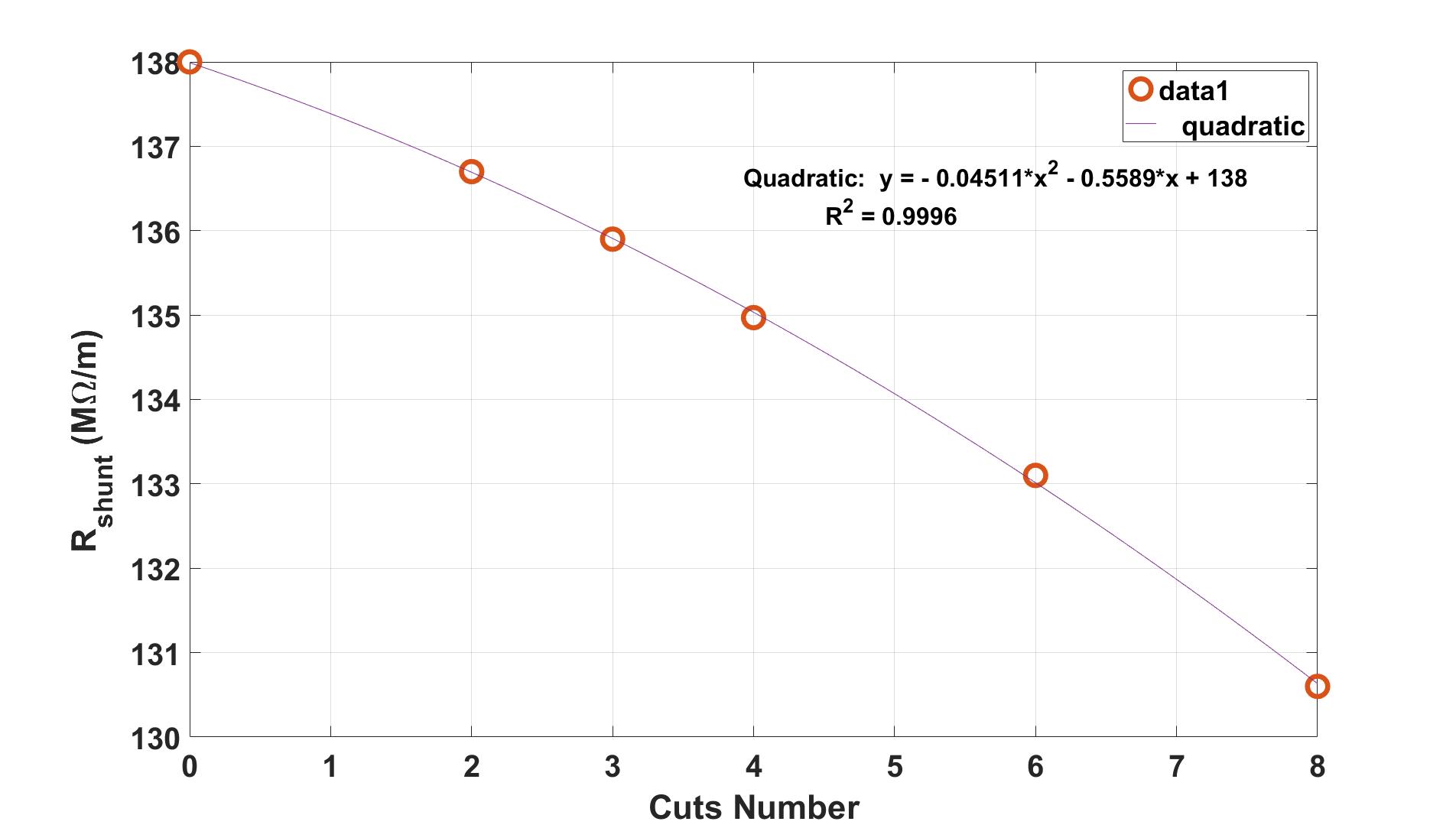}
\caption{\label{Rshunt_new}Longitudinal shunt impedance $R_{sh}$ of the operating mode as function of the number of cuts.}
\end{figure}

\begin{figure}[ht]
\centering
\includegraphics[width=1.1\columnwidth]{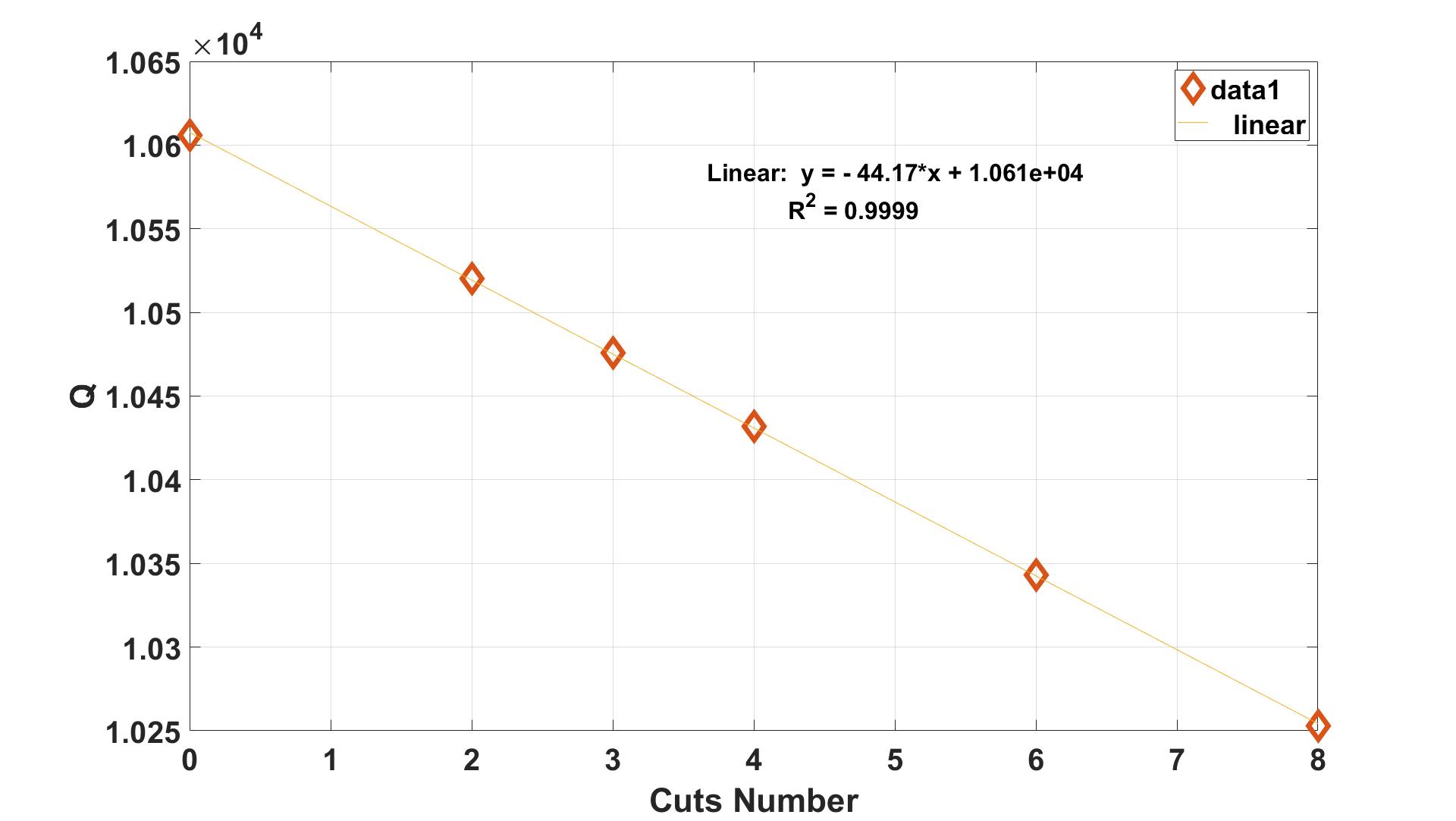}
\caption{\label{Q_new} Quality factor Q of the operating mode as function of the number of cuts.}
\end{figure}

Figures \ref{Rshunt_new} and \ref{Q_new} show the longitudinal shunt impedance $R_{sh}$ and quality factor $Q$ of the operating mode as function of the number of cuts (coinciding with the number of sectors). As a result, the  perturbation introduced with a 1 mm gap region affects slightly the  operating mode longitudinal shunt impedance $R_{sh}$ and the quality factor $Q$. The quality factor reduction as a function of the quadrants number is clearly linear. Its fit is estimated to be $Q \simeq  - 44.17 n + 10610$, and it has a  linear behaviour with a change rate of about $dQ/dn  \simeq - 44.17$.  The fit of the longitudinal shunt impedance $R_{sh}$ dependence as a function of the quadrants number is instead quadratic: $R_{sh} \simeq - 0.04511 n^2 – 0.5589 n + 138$.

With the structure made up of 8 sectors, the total quality factor reduction is estimated to be of about  0.3 \% while the longitudinal shunt impedance reduction $R_{sh}$ of the operating mode is of about 0.6 \%, both absolutely negligible. Both $Q$ and $R_{sh}$ variations are physically consistent, and they are in agreement with the well known and basic definitions of the cavity merit factor $Q = \omega U/P$  and $R_{sh} = V_c^2/P$ where $U$ is the cavity stored energy, $P$ the power dissipation and $V_c$ the integrated voltage on axis since the gap regions affect slightly the frequency of the operating $\pi$ mode, yield  a proportional reduction of $U/P$ ratio and cause a quadratic dependence of the $R_{sh}$ decrease.

\begin{figure}[ht]
\centering
\includegraphics[width=1.1\columnwidth]{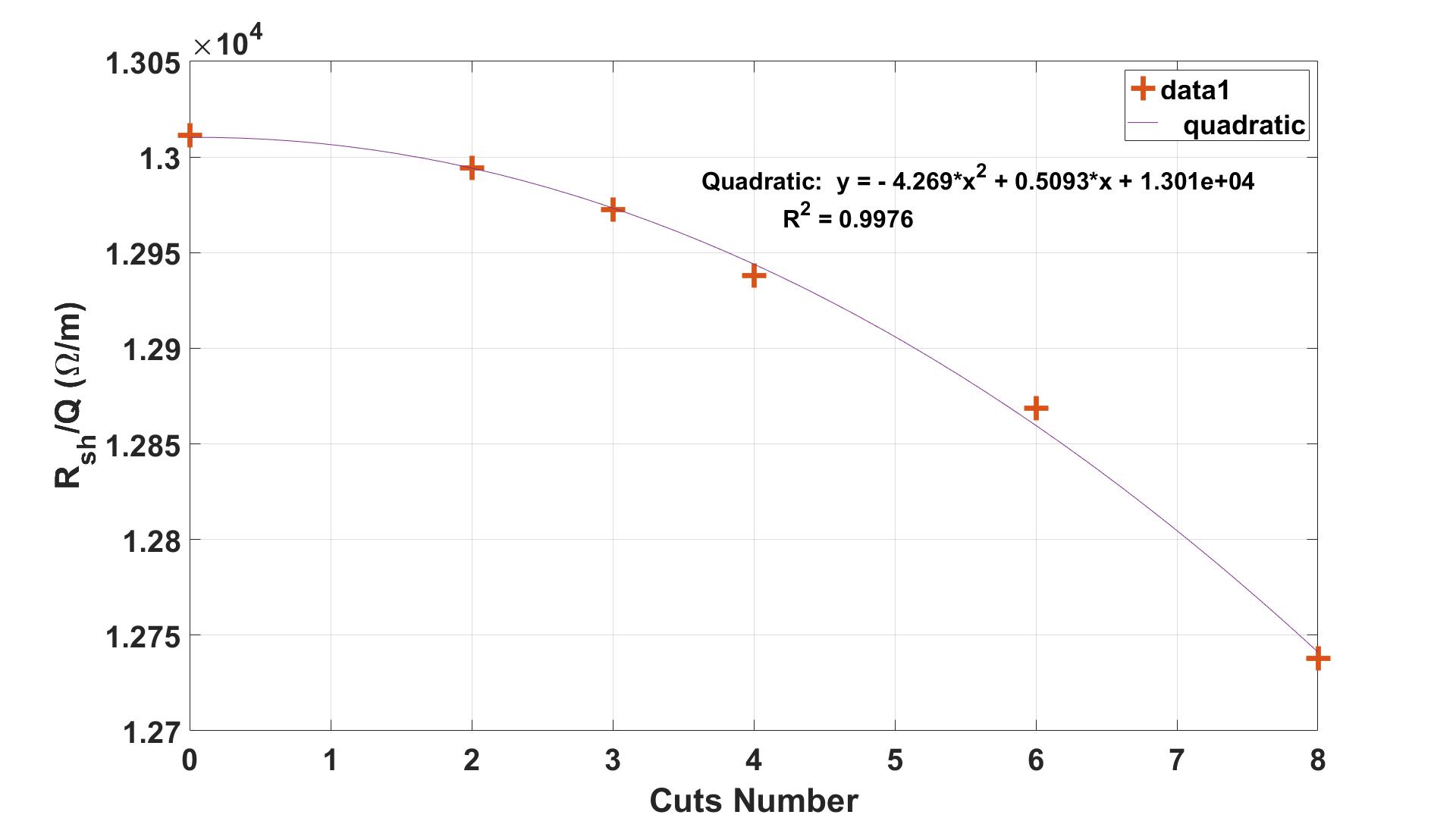}
\caption{\label{R_Q_new} Form factor $R_{sh} /Q$ of the operating mode as function of the number of cuts by ignoring  the beam transit time.}
\end{figure}

Figure  \ref{R_Q_new} illustrates the form factor $R_{sh} /Q$ of the operating mode as function of the gap regions by ignoring  the beam transit time. The fit of the form factor as a function of the number of sectors is also quadratic: $R_{sh}/Q \simeq - 4.269 n^2 + 0.5093 n + 13010$.

\begin{figure}[ht]
\centering
\includegraphics[width=1.1\columnwidth]{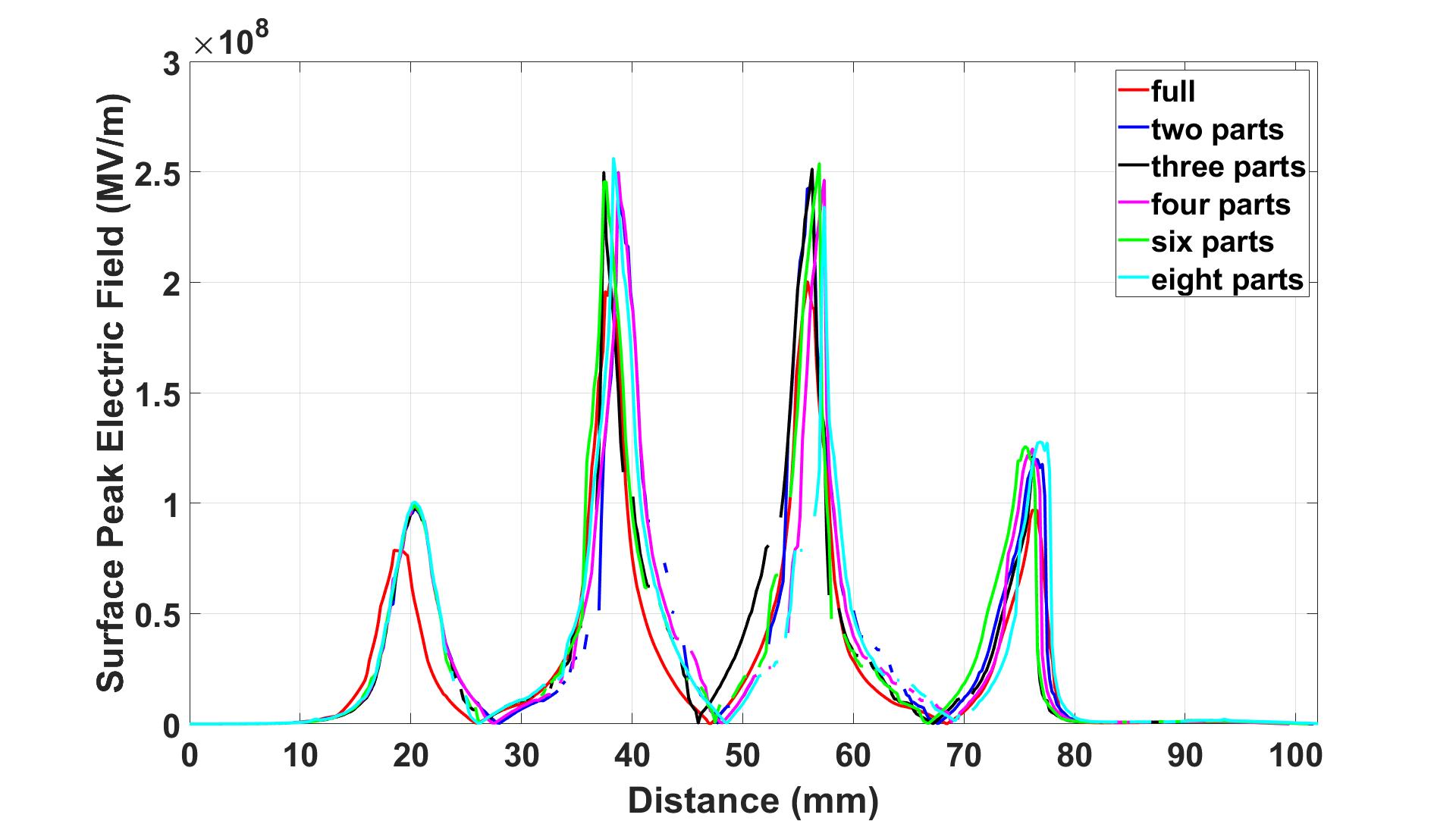}
\caption{\label{Epeak_new} Surface peak electric field plotted along the cavity edge for all cases of gap numbers.}
\end{figure}

As additional investigation, we estimated the electric and modified Poynting vector field profiles along the edge close to the structure cut. Figure \ref{Epeak_new}  shows the  surface peak electric field plotted along the cavity edge for different number of sectors. The field profile shows similar peak values around the middle cell for all cases. These values are about 20\% lower than the values reported in Table \ref{tab:Cavity_RF_Parameters} that refer to the locations of highest curvature of the fillet between cavity edge and the cut.

\begin{figure}[ht]
\centering
\includegraphics[width=1.1\columnwidth]{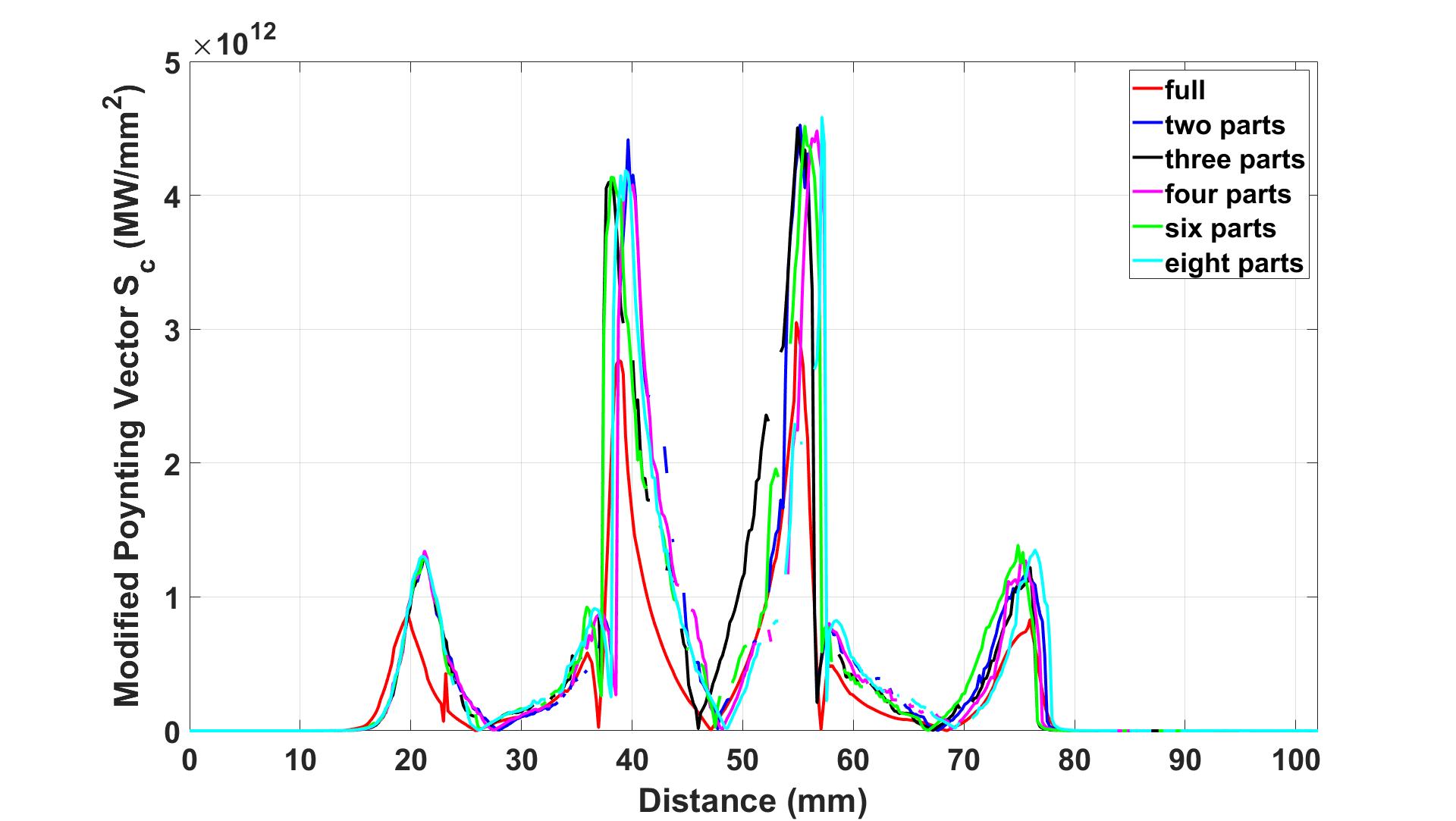}
\caption{\label{Sc_edge_new} Poynting vector $S_c$ plotted along the cavity edge for different number of sectors.}
\end{figure}

The same behaviour occurs for modified Poynting vector $S_c$ shown in Figure \ref{Sc_edge_new}  which is plotted along the cavity edge for the different studied cases.

\section{Experimental setup}

A prototype, consisting in two halves, has been realised by COMEB \cite{ref12}. The two cavity halves are aligned and clamped together by means of male-female matching surface. The clamping is obtained with stainless screws, and the cavity will be TIG welded at COMEB on the outer surface. 

The structure has been realized by mechanical machining with a numerically controlled milling machine and the obtained precision is about  $\pm 2 \mu$m, while the surface roughness is not worse than 0.3 $\mu$Ra. The surface finishing was obtained directly by mechanical machining with custom cutting tools, avoiding any polishing technique and only silicon and sulphur free cutting fluid was used. The final machining was done at constant temperature in order to guarantee as much as possible the uniformity of the mechanical dimension of the cells. After machining, a standard cleaning procedure was performed using an alkaline solution at 3\% at 50$^\circ$ Celsius. This was followed by a rinse first in tap water and then in distilled water. Then a chemical cleaning with citric acid solution was performed with additional rinses. Finally the pieces were dried in a dust free oven. Each cell dimension has been checked with a quality control test. 

\begin{figure}[hb!]
\centering
\includegraphics[width=1\columnwidth]{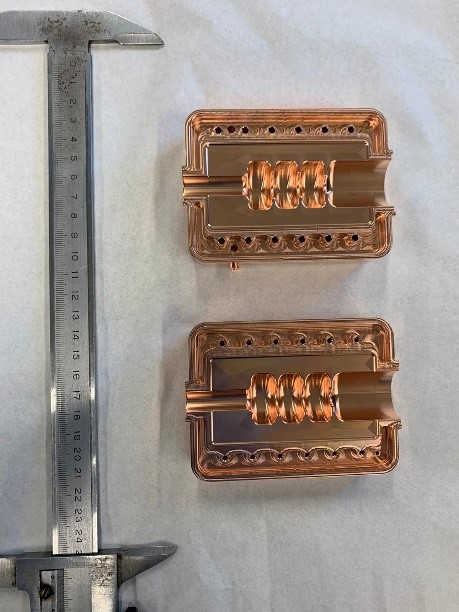}
\caption{\label{Pre_TIG} Two structure halves after machining.}
\end{figure}

\begin{figure}[hb!]
\centering
\includegraphics[width=1\columnwidth]{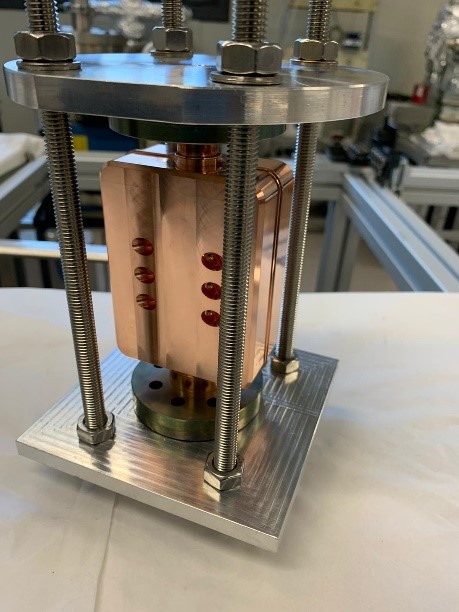}
\caption{\label{Pre_TIG2}The clamped split cavity before TIG welding.}
\end{figure}

\begin{figure}[h!]
\centering
\includegraphics[width=1.2\columnwidth,angle=-90]{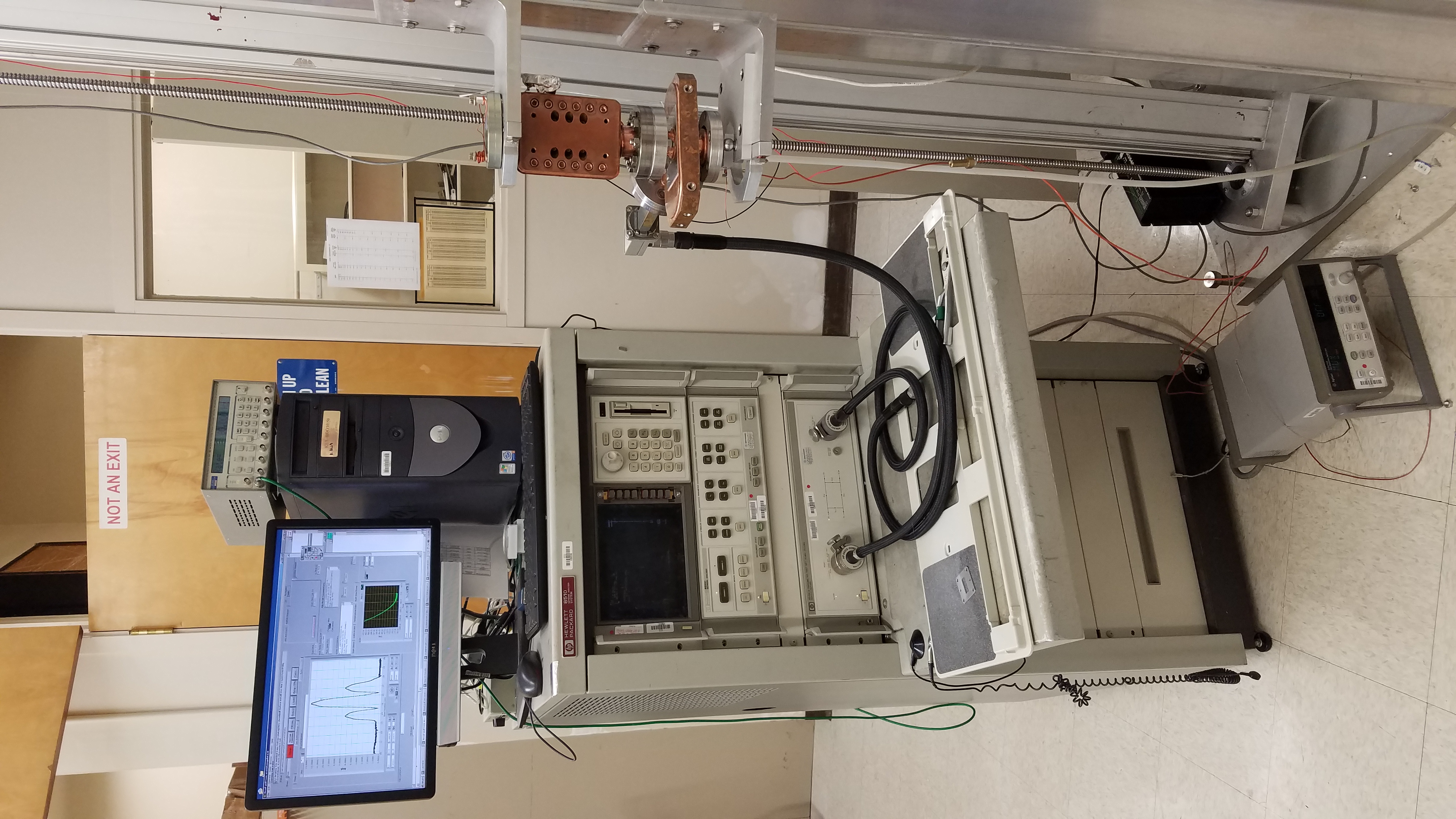}
\caption{\label{Setup}Complete measurement setup.}
\end{figure}

In Figure \ref{Pre_TIG}, we show the two structure halves after machining. The clamped split cavity before TIG welding is given in Figure \ref{Pre_TIG2}. The complete measurement setup is shown in Figure \ref{Setup}. The PC controls both the network analyser Agilent N5230A (interfaced by a GPIB Ethernet device) and the control circuit of the steeping motor through LABVIEW \cite{ref44}. The nylon wire is kept straight by a 75 g weight. The initial low level RF characterisation have been performed at SLAC. 

\begin{figure}[h!]
\centering
\includegraphics[width=1\columnwidth]{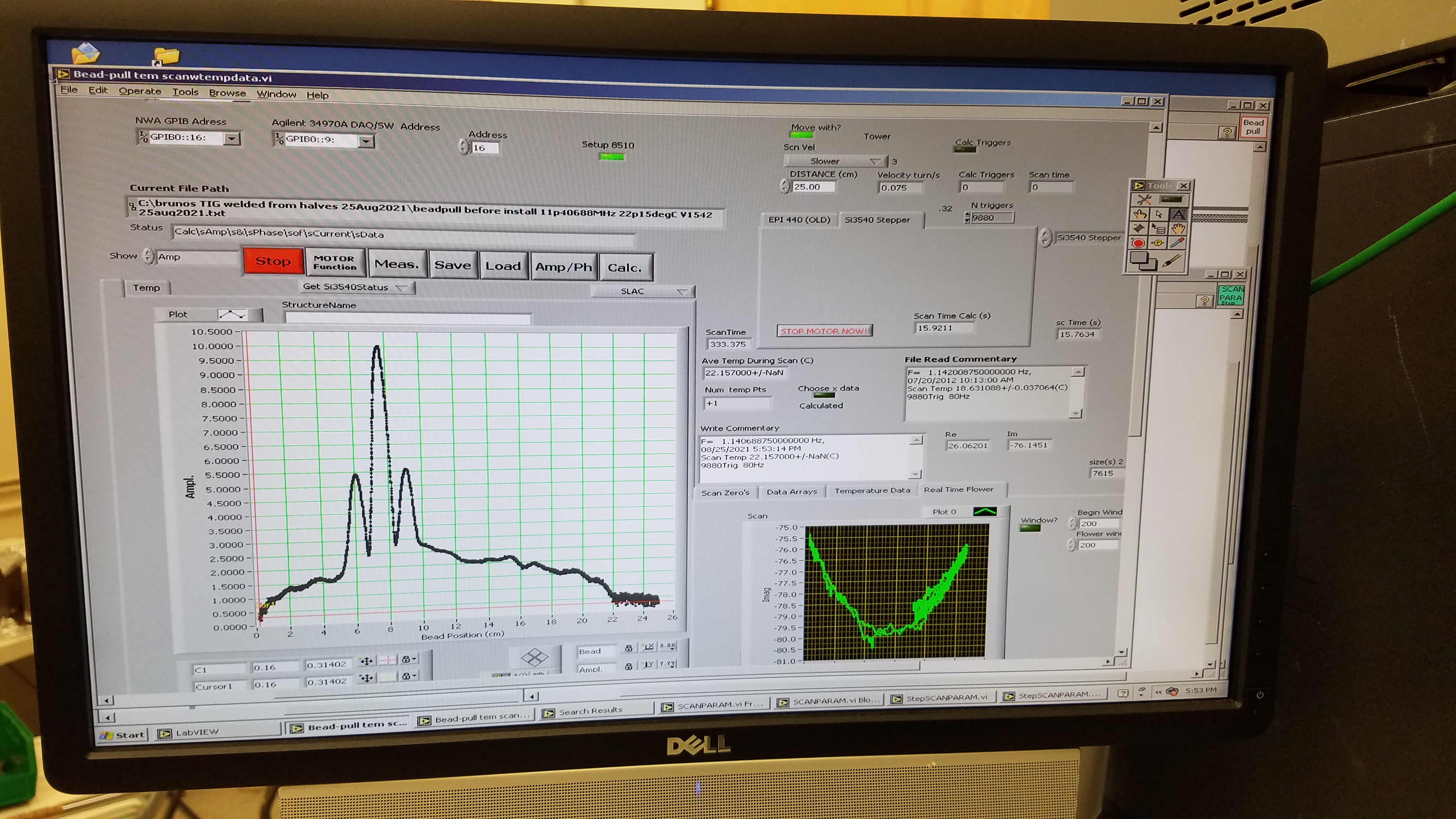}
\caption{\label{Labview}LabView plot of the on-axis electric field in the clamped cavity measured with the bead-pull technique.}
\end{figure}

Figure \ref{Labview} shows a LabView plot of the on-axis electric field in the clamped cavity measured with the bead-pull technique. The measured field profile is in a good agreement with the simulated one. The measured resonant frequency of the operating $\pi$ mode of the nitrogen-filled cavity was 11.4186 GHz with a good quality factor of 9500. 

Two types of different measurements have been carried out:  transmission (or reflection) scattering coefficients and bead pull measurements. With the first type of measurements we have found the resonant frequency, the reflection coefficient at the input ports and the unloaded or external quality factors of $\pi$ (or other) mode(s). With the second type of measurements we have found the longitudinal electric field on axis and we have calculated the shunt impedance of the structure.

The measured resonant frequency of the operating $\pi$ mode corrected to vacuum was 11.4193 GHz. The measured resonant frequency of the operating $\pi$  mode corrected to vacuum was 11.4193. This particular structure is well within SLAC klystron bandwidth.

\section{Applications of the Split Open Structures}

The next generation of accelerators is highly demanding in terms of maximizing accelerating gradients, minimizing overall machine length and cost, improving the beam quality, reducing beam loading effects and so on. In the case of circular accelerators one important  mechanism that produces beam instability and power losses consists in the possibility of the charged particles to excite high-order resonant fields in RF accelerating cavities when passing through them. This is possible only for those accelerating structures with resonant modes in frequency ranges that overlap the beam Fourier spectrum. If the beam current is high, coupled-bunch instabilities may occur. 

Furthermore, in pre-buncher and chopper cavities of linear accelerators operating at high currents, an RF detuning is required to reduce the effects of the interaction beam-structure, which could affect the beam quality. These problems can be cured by using the multi-sector structure approach, which also provides the possibility of improving the vacuum level of a factor of at least ten with respect to the usual solution, by inserting an additional vacuum chamber connected to the beam pipe. Additionally, in recent years, we have seen a revolution in high-brightness electron beams because of the maturation of RF photo-injector performance. In order to improve the electron beam brightness, larger mode separation is also needed. For a fixed geometry of the structure (which means a fixed form factor ratio $R_{sh}/Q$) and for a given beam current and cavity peak voltage, the frequency detuning effect is determined. 

We conjecture that, with further development, welded structures of this type can be used as simple practical accelerators by adding more cells and couplers.  As an example, we have already observed in the X-band lineariser for the Coherent Light Source at INFN-LNF \cite{ref24} the overlapping  between the operating $\pi$ mode and the nearest one 8/9 $\pi$ of the RF structure passband. To solve this problem it was  decided to feed the cavity in the central cell in order to not excite the mode 8/9 $\pi$. So, with a central coupler we were able to get a much greater separation of the mode frequencies since the working mode is less perturbed by the closest one, by allowing to achieve  a stable operation. However, by increasing the number of cells, the overlapping between the operation mode and and nearest  one becomes even more important and probable. As a result, a structure fabricated with sectors, provides a great advantage in terms of  mode frequency separation by maintaining, at the same time, a high longitudinal shunt impedance of the working mode.

In a linear  accelerating structure constructed with sectors, a lossy material can be installed in gaps of cavity to damp HOMs. Then whole body can be TIG or electron beam welded for making a hard structure. High power tests will be performed with the Slac’s lead box.  Moreover, the detuning of dipole mode could be natural if  we can use constant gradient TW structures. However, the combination of damping  and detuning and TW parallel coupling is a novel technique. Intense investigations on the absorbers for damping the higher order modes, are also in progress. In addition, we are currently planning to use the possibility of cryogenic accelerating structures since preliminary tested  cryogenic X-band cavity  provided  a surface electric field of about 500 MV/m or 250 MV/m accelerating gradient for  a shaped RF pulse of 150 ns of flat gradient with a rf breakdown rate of $2\times10^{-4}$/pulse/m. This approach is also very attractive for a RF photoinjector  \cite{ref0.1}, where a very large gradient would lead to very small emittances.

\section{Conclusions}

The technological activity of testing high-gradient RF sections is related to the investigation of breakdown mechanisms, which limit the high gradient performance of these structures. Since experimental results with hard copper cavities, conducted at SLAC, CERN
and KEK  have shown that hard materials sustain higher accelerating gradients for the same breakdown rate, we have presented multi-sector X-band structure made up with hard Cu/Ag alloy, assembled by using the TIG technology which allows to increase the frequency separation of the passband mode. The realised prototype will be used to investigate the RF breakdown physics. Our interest is to show the viability of  this method for future accelerators. Installation of absorbers for damping the HOM is also planned.

In addition, multiquadrants accelerating structures can be utilized as radiofrequency photoinjectors for FEL and Compton sources in single-bunch operation since we conjecture that the beam quality improves due to the mode frequencies separation. Dedicated beam dynamics simulations will be performed according to the specific application.

Due to the Meissner shielding currents, the penetration of the electromagnetic field in a superconductor is confined to a very thin superficial layer. For instance, a 1 $\mu$m thick Nb film is largely sufficient to shield completely the underlying more resistive copper. For this reason,  in the RF superconducvity field , the Cu-Nb alloy is  preferred and more promising  than  the conventional Nb bulk. In addition, potentially, the accelerating cavities made with quadrants and Copper coated  Nb, can be used for the RF superconducting accelerators.

In the framework of the UC-XFEL project to be constructed at UCLA, a 34 GHz standing wave accelerating structure has been designed for the phase-space linearization of electron beams. The required input RF power for two cavities, with an 8 cm length each, is 13.2 MW. We have foreseen two options for the RF power source. The first one, in normalconducting operation, is the use of the magnicon which can provide up to 40–50 MW with a pulse compressor. The second one, in cryogenics operation at 77 K, uses a gyroklystron which can provide up to 12 MW with the use of an RF pulse compressor. The two SW structure solution provides an integrated voltage of 16 MV and it avoids the use of a high-power circulator [5 and therein].

As a final comment, we intend to apply the same construction method for the Ka-band structures in the framework of the CompactLight and UC-XFEL projects.

\begin{acknowledgments}
This work was partially supported by INFN National committee V through the DEMETRA and ARYA projects.
\end{acknowledgments}

\end{document}